# Cooperative effects in molecular conduction

Arie Landau[1], Leeor Kronik[2] and Abraham Nitzan[1†]

[1] School of Chemistry, the Sackler Faculty of Science, Tel Aviv University, Tel Aviv, 69978, Israel
[2] Department of materials and interfaces, the Weizmann Institute of Science, Rehovoth, 76100, Israel

## Abstract

Current experimental and theoretical studies on the effect of intermolecular interactions on molecular conduction appear to be in conflict with each other. In particular, some experimental results, e.g., the observation of 2-dimensional free-particle character for interface bound electrons indicate strong intermolecular interactions while other observations indicate an additive character of conduction properties. In this paper we use a generic tight binding model with a physically motivated choice of parameters in order to examine this issue. The model encompasses direct intermolecular interactions as well as through-metal interactions and can be solved exactly to yield spectral properties (surface density of states) and transport characteristics (transmission coefficients and current-voltage behaviors) for single molecule junctions, molecular islands and molecular layers. We find linear scaling of conduction properties with the number of conducting molecules in junctions characterized by molecular layers when the probe (STM tip) addresses different numbers of molecules; however, the conduction per molecule can differ significantly from that of an isolated single molecule. When a junction involves finite molecular islands of varying sizes, linear scaling sets in only beyond a certain molecular island size, of the order of a few tens of molecules. Implications for current observation of linear scaling behaviors are discussed.

† Corresponding Author: Nitzan@post.tau.ac.il

## 1. Introduction

The electronic conduction properties of metal-molecule-metal junctions are determined by the electronic properties of the metal and molecular constituents, the bonding between them, the junction structure and configuration, external electrostatic (gate) fields and environmental parameters such as temperature. The relationship of function and structure offers routes for controlling the junction operation but also results in uncertainties about performance and stability. Characterization of such relationships is therefore central to the study of molecular conduction junctions. One such issue that was already addressed in several experimental and theoretical papers is dependence on the number of molecules involved in the conduction process. In particular, a comparison between electronic transport of a single molecule junction and that of a junction comprising the corresponding molecular monolayer is of interest. While a common practice is to assume linear scaling with molecular coverage, i.e. to associate the conduction-per-molecule observed in monolayer junctions with that of the single molecule junction, direct experimental and numerical examinations yield mixed observations on this point. On the one hand clear evidence of linear scaling is indicated by the results of Cui et al[1] and Xu and Tao[2] in junctions containing 1-5 n-alkane molecules.[1] Similarly Kushmerik and coworkers[3] have found that the current-voltage curves of monolayers of isonitrile oligo(phenylene ethynylene) molecules measured in the cross wire technique with varying contact areas can be scaled to a single curve by dividing with different integers in the range 1-1000, suggesting that the scaled curve corresponds to a single molecule junction. Supporting evidence is obtained by comparing STM and cross wire conduction measurements on a series of saturated and conjugated molecules,[4] where similar current-voltage traces where obtained by using a multiplication factor (of the order of $10^3$) that may account for the different number of molecules in these junctions. On the other hand, the conduction measured in some single molecule junctions was found to be several orders of magnitudes larger than the conduction-per-molecule mentioned in the corresponding monolayers.[5] In a more recent study Selzer et al[6] have noticed that the conduction per molecule of a molecular layer is similar to that of the corresponding single molecule, however the differential conductance of the latter increases considerably more rapidly with bias and can become a thousand time larger than the per-molecule layer conductance. Furthermore, the single molecule (1-nitro-2,5-di(phenylethynyl-4'-mercapto)benzene) junction shows temperature dependence with transition to activated conduction at $T > 100K$, while the conduction of the corresponding molecular layer remains temperature independent up to room temperature. Finally, a recent scanning tunneling spectroscopy study[7] of monolayers of 3,4,9,10-perylenetetracarboxylicacid- dianhydride (PTCDA) on a silver substrate showed a remarkable electronic band dispersion associated with the layer electronic structure, indicating a free electron behavior with an effective mass of $m_{eff} = 0.47 m_e$ ($m_e$ is the free electron mass). This surface electron delocalization indicates strong intermolecular coupling that has been attributed mostly to through-metal interaction. The existence of such coupling stands in contrast to assertions that adsorbed monolayer behave as collections of non-interacting molecules, and suggests that conduction in this system will not scale linearly with that of single-molecule junctions based on the same molecule.

---

[1] In Ref. [2] also 4,4' bipyridine was used with similar results. Note that the conduction values reported by Xu and Tao in these papers are about an order of magnitude larger than those of Cui et al; the difference probably resulting from unrelated technical factors.

Theoretical studies of this issue also lead to varied results. Early studies by Magoga and Joachim[8] conclude that linear scaling of the conduction $g(N)$ with the number $N$ of identical molecules connecting the leads exists in the form

$$g(N) = N g_{eff}(1), \qquad (1)$$

where the effective single molecule conduction $g_{eff}(1)$ differ from the corresponding single molecule property $g(1)$ because of intermolecular interactions. The latter can be direct interactions between neighboring molecules or interactions resulting from their mutual coupling to the leads. Obviously however, $N$ has to be large enough for Eq. (1) to hold. In the non-resonant small bias regime $g_{eff}(1) > g(1)$, because intermolecular interaction brings molecular states density closer to the Fermi energy of the lead. Yaliraki and Ratner[9] have made a similar observation by comparing the conduction of two parallel molecular wires to that of a single wire. On the other hand, Lang and Avouris[10] have studied the low bias conduction of a junction containing two carbon wires connecting jellium leads as a function of the interwire separation. A non-monotonic distance dependence, with an overall conduction increase at larger separation was found and rationalized in terms of the effect of the interwire interaction on the molecular density of states at the Fermi energy. The magnitude of the effect should depend on the adsorbate molecules. Indeed, Kim[11] et al have studied the conduction of junctions based on single molecules and molecular layers of Hexane-dithiolate between Au(111) electrodes, and found that the effect of intermolecular coupling on conduction is negligible in these junctions. Still, polar adsorbate molecules, or molecules that form polarized adsorption sites because of charge transfer to/from the substrate are expected to show strong intermolecular interaction effects.[12-16] For example, Deutsch et al[14] have found that the charge transfer characteristic of single benzene derivative molecules adsorbed on Si(111) are reversed relative to the corresponding molecular layer, as a result of the dipole-dipole interaction in the latter.

Effects of intermolecular interactions on properties of molecular junctions can arise from several sources. Clearly, direct interactions between the molecules may affect the transport properties of a molecular layer, as can indirect interactions mediated by the underlying substrate. Such interactions may directly affect transport, but may also influence transport properties by affecting the molecular configuration and its response to the imposed bias potential and to temperature change. Electrostatic effects are particularly important. In addition to affecting charge transfer properties as discussed above, it has been shown[17,18] that the way an imposed bias is distributed across the junction depends on the lateral dimension of the bridging layer in a way that reflects the layer screening properties. Finally, junction stability is related to its structural response to the imposed field and ensuing current as well as to its temperature rise during conduction. The latter depends on the junction ability to dissipate excess energy which in turn depends on its immediate environment. In particular large differences may exist between a junction that can dissipate heat only through the contacting electrodes (as with a single molecule connecting metal leads in vacuum) or also through lateral interactions with neighboring molecules (a situation associated with molecular layers and junctions embedded in solvent environments).

This paper focuses on one aspect of single molecule versus molecular layer behavior by presenting a generic study of effects associated with short range intermolecular and molecule-substrate interactions on the transport properties of single molecules and the corresponding molecular islands and layers. We use a minimal model of single-level molecules connecting electrodes characterized by simple cubic lattices; such a model, characterized by a physically motivated choice of coupling parameters, appears to capture all the essential physics of this problem. We examine structures of three types commonly encountered in experimental systems: single molecules, adsorbed layers of commensurate and incommensurate structures, and single molecules embedded in a molecular island or layer. It should be pointed out that some experimental reports of single molecule behavior are indeed associated with structures of the latter type.

Section 2 describes our model and theoretical approach. Section 3 describe the calculation of the required Green functions and self energies. Section 4 formulate the spectral and transport properties and the ensuing current voltage characteristics of different junction structures associated with single molecules, molecular islands and molecular layers. Section 5 discusses the chosen model parameters and in section 6 presents some numerical results that compare physical observables for different junction structures associated with the same interaction parameters. Section 7 concludes.

## 2. Model

The systems investigated here are a molecular monolayer (ML), a finite molecular island (IL) and a single molecule (SM) chemisorbed on the surface of a tight binding metallic model. We examine the difference in the transmission probability through a ML, IL and SM connecting two metallic electrodes. In addition, a finite molecular island system is examined. The metal electrode is described by a simple-cubic cell structure characterized by a lattice constant, $a$ (Fig. 1). This lattice is semi-infinite in the $Z$ direction perpendicular to the metal surface and periodic boundary conditions are used in the X and Y directions. A lattice point is described by $(n_x, n_y, n_z)$ where

$$n_x = -N_x/2, -(N_x/2)+1, ..., (N_x/2)-1,$$
$$n_y = -N_y/2, -(N_y/2)+1, ..., (N_y/2)-1 \qquad (2)$$
$$n_z = 0, 1, ..., \infty,$$

the metal surface is indexed as $n_z=1$ and The molecules are taken to occupy positions on the plane $Z=0$; for the simplest commensurate molecular layer this positions corresponds to all lattice sites. Both the metal and the molecular system are described by simple nearest-neighbor (nn) tight-biding models. Each metallic-atom is represented by a single orbital $|n_x,n_y,n_z\rangle$, coupled to its nearest neighbor orbitals. A similar description is also used for the molecular system and for the molecule–metal interaction. The model is characterized by the molecular and metal site energies, $\varepsilon_m$ and $\varepsilon_t$, respectively, and by the interaction between neighboring metal sites, $V^{(t)}$, neighboring molecules, $V^{(m)}$, and nearest-neighbor metal-molecule interaction $V^{(mt)}(n_x,n_y)$. The model Hamiltonian $\hat{H} = \hat{H}^{(m)} + \hat{H}^{(t)} + \hat{H}^{(mt)}$ has the usual nn tight binding form where, e.g., the molecule-metal interaction is given by

$$\hat{H}^{(mt)} = V^{(mt)} \sum_{n_x}\sum_{n_y} \left[ |n_x n_y (n_z = 0)\rangle\langle n_x n_y (n_z = 1)| + h.c. \right]. \qquad (3)$$

The sum over the lateral indices $n_x$, $n_y$ is over the occupied sites of the molecular adsorbate, Z=0, plane. More than one coupling parameter $V^{(mt)}$ may be needed to describe more complex layer structures (see below).

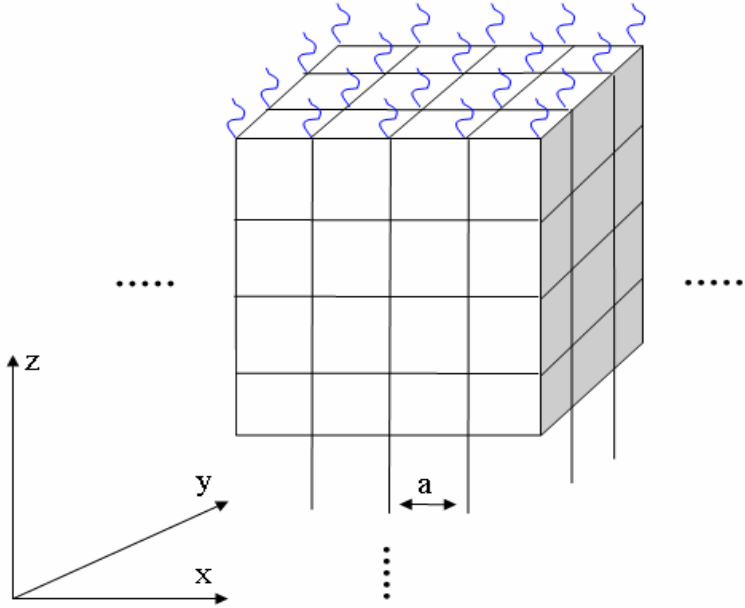

Fig. 1. Schematic cartoon of a simple 1:1 ML (curly lines represent molecules) adsorbed on a metallic electrode (grid points represent metallic-atoms). The Z direction is normal to the surface, X and Y are parallel to it, *a* is the unit distance.

It is convenient to replace the local basis $|n_x,n_y,n_z\rangle$ by a basis in which Bloch wavefunctions are used in the X, Y directions:

$$\left|\theta_x\theta_y n_z\right\rangle = \frac{1}{\sqrt{N_x N_y}} \sum_{n_x=-(N_x/2)}^{(N_x/2)-1} \sum_{n_y=-(N_y/2)}^{(N_y/2)-1} e^{i(\theta_x n_x + \theta_y n_y)} \left|n_x n_y n_z\right\rangle, \quad (4)$$

where $\theta_u = k_u a$, and $k_u = 2\pi j / N_u$ $(j = 1, 2, ..., N_u; u = x, y)$.

In this representation the metal Hamiltonian takes the form

$$\mathbf{H}^{(t)} = \begin{pmatrix} \mathbf{H}^{(t)}_{\boldsymbol{\theta}_{xy}} & 0 & 0 & \cdots \\ 0 & \mathbf{H}^{(t)}_{\boldsymbol{\theta}'_{xy}} & 0 & \cdots \\ 0 & 0 & \mathbf{H}^{(t)}_{\boldsymbol{\theta}''_{xy}} & \cdots \\ \cdots & \cdots & \cdots & \cdots \end{pmatrix} \quad (5)$$

where we have denoted $\boldsymbol{\theta}_{xy} = (\theta_x, \theta_y)$ and where each block on the diagonal represents a 1-dimensional problem in the Z direction and has the form

$$\mathbf{H}^{(t)}_{\boldsymbol{\theta}_{xy}} = \begin{pmatrix} E_t(\boldsymbol{\theta}_{xy}) & V^{(t)} & 0 & \cdots \\ V^{(t)} & E_t(\boldsymbol{\theta}_{xy}) & V^{(t)} & \cdots \\ 0 & V^{(t)} & E_t(\boldsymbol{\theta}_{xy}) & \cdots \\ \cdots & \cdots & \cdots & \cdots \end{pmatrix} \quad (6)$$

where

$$E_t(\boldsymbol{\theta}_{xy}) = \varepsilon_t + 2V^{(t)}(\cos\theta_x + \cos\theta_y) \quad (7)$$

## 3. Green functions and spectral properties

Consider first the bare metal. From (5) it follows that the retarded metal Green's function (GF) is also block diagonal, $G^{(t)}_{\boldsymbol{\theta}_{xy}}(E) = \lim_{\eta \to 0} \left(E - H^{(t)}_{\boldsymbol{\theta}_{xy}} + i\eta\right)^{-1}$. A standard method for evaluating the surface GF is to use the semi-infinite periodic structure of the metal along the Z direction.[19] Because the block Hamiltonian, Eq. (6), has the form

$$\mathbf{H}^{(t)}_{\boldsymbol{\theta}_{xy}} = \begin{pmatrix} E_t(\boldsymbol{\theta}_{xy}) & V^{(t)}_t \\ V^{(t)} & \mathbf{H}^{(t)}_{\boldsymbol{\theta}_{xy}} \end{pmatrix} \quad (8)$$

It follows that

$$G^{(ts)}_{\theta_{xy}}(E) = \left(E - E_t(\theta_{xy}) - \Sigma^{(ts)}_{\theta_{xy}}(E)\right)^{-1} \tag{9}$$

where the surface GF, $G^{(ts)}_{\theta_{xy}}(E)$, is the ($n_z = 1$, $n_z = 1$) element of the matrix $G^{(t)}_{\theta_{xy}}$ and where

$$\Sigma^{(ts)}_{\theta_{xy}} = V^{(t)} G^{(ts)}_{\theta_{xy}} V^{(t)} \tag{10}$$

is the corresponding self energy (SE). Eqs. (9) and (10) can be solved for the metal surface GF iteratively. In addition, the surface spectral density function of the metal is defined as

$$\Gamma^{(ts)}_{\theta_{xy}}(E) = -2\,\mathrm{Im}\left[\Sigma^{(ts)}_{\theta_{xy}}(E)\right] \tag{11}$$

Next consider the case of a simple[2] adsorbed monolayer as shown by Fig. 1. The above derivation can now be repeated for this layer as a surface layer; the difference enters only through the different energy parameters. Consequently,

$$G^{(ml)}_{\theta_{xy}}(E) = \left(E - E_m(\theta_{xy}) - \Sigma^{(ml)}_{\theta_{xy}}(E)\right)^{-1} \tag{12}$$

where

$$E_m(\theta_{xy}) = \varepsilon_m + 2V^{(m)}\left(\cos\theta_x + \cos\theta_y\right) \tag{13}$$

and

$$\Sigma^{(ml)}_{\theta_{xy}} = V^{(mt)} G^{(ts)}_{\theta_{xy}} V^{(mt)} \quad ; \quad \Gamma^{(ml)}_{\theta_{xy}}(E) = -2\,\mathrm{Im}\left[\Sigma^{(ml)}_{\theta_{xy}}(E)\right] \tag{14}$$

When the molecular layer is in contact with two metals, L and R, with the same simple 1:1 adsorption geometry, the ML SEs combine additively, $\Sigma^{(ml)}_{\theta_{xy}} = \Sigma^{(ml,L)}_{\theta_{xy}} + \Sigma^{(ml,R)}_{\theta_{xy}}$, where

$$\Sigma^{(ml,K)}_{\theta_{xy}} = V^{(mt,K)} G^{(ts,K)}_{\theta_{xy}} V^{(mt,K)}; \quad K = L, R. \tag{15}$$

In what follows we also require the position space (XY) representation of these functions. These are obtained from

$$G^{(ts)}_{\mathbf{n}_{xy},\mathbf{n}_{xy}'}(E) = \sum_{\theta_x}\sum_{\theta_y}\langle \mathbf{n}_{xy}|\boldsymbol{\theta}_{xy}\rangle\langle \boldsymbol{\theta}_{xy}|\mathbf{n}_{xy}'\rangle G^{(ts)}_{\boldsymbol{\theta}_{xy}}(E)$$
$$= \frac{1}{4\pi^2}\int_0^{2\pi}\int_0^{2\pi}d\theta_x d\theta_y G^{(ts)}_{\boldsymbol{\theta}_{xy}}(E) e^{i(\boldsymbol{\theta}_{xy}\cdot\Delta\mathbf{n}_{xy})} \quad (16)$$

where $\mathbf{n}_{xy}\equiv(n_x,n_y)$ and $\Delta\mathbf{n}_{xy}=(n_x-n_x',n_y-n_y')$. Similar transformations hold for other functions of $\boldsymbol{\theta}_{xy}$. In particular the local site GF

$$G^{(ml)}_{\mathbf{n}_{xy},\mathbf{n}_{xy}}(E) = \frac{1}{4\pi^2}\int_0^{2\pi}\int_0^{2\pi}d\theta_x d\theta_y G^{(ml)}_{\boldsymbol{\theta}_{xy}}(E), \quad (17)$$

is related to the local density of states (DOS) per molecule

$$\rho^{(ml)}(E) = -\frac{1}{\pi}\text{Im}\left[G^{(ml)}_{\mathbf{n}_{xy},\mathbf{n}_{xy}}(E)\right]. \quad (18)$$

---

[2] By "simple" we mean that each adsorbed molecule is coupled to only one metal surface atom, and vice verse.

Similar expressions can be obtained for the metallic surface DOS using the surface GF (16).

Consider next the case where instead of a homogeneous molecular layer we have a molecular island of $J$ molecules adsorbed on the metal surface and interacting with $N$ surface atoms. Denote the corresponding interaction elements by $V^{(mt)}_{nj}$; $n=1,..,N$; $j=1,..,J$. Note that there is a one-to-one correspondence between $n$ and a position $\mathbf{n}_{xy}$ on the metal surface, however we have dropped the requirement of "simplicity" as defined above[2] with respect to the adsorbate island or ML. The molecular island Hamiltonian, $\hat{H}^{(il)}$, is a combination of a diagonal part $\varepsilon_m \mathbf{1}_J$ where $\mathbf{1}_J$ is a unit matrix of order $J$, and a non-diagonal part associated with the interactions $V^{(m)}$ between island molecules. The island GF (a $J\times J$ matrix in the position representation) is now given by

$$G^{(il)}(E) = \left(E - \hat{H}^{(il)} - \hat{\Sigma}^{(il)}\right)^{-1} \quad (19)$$

where the island SE is given by

$$\Sigma^{(il)}_{j,j'} = \sum_n^N\sum_{n'}^N V^{(mt)}_{jn}G^{(ts)}_{n,n'}V^{(mt)}_{n'j'}; \quad \Gamma^{(il)}_{j,j'}(E) = i\left(\Sigma^{(il)} - \Sigma^{(il)\dagger}\right)_{j,j'} \quad (20)$$

and may be easily evaluated for any finite island once the metal surface GF (Eq. (16)) has been calculated. Again, when the molecular island connects between two metals the self-energy in (19) is the sum $\Sigma^{(il,L)} + \Sigma^{(il,R)}$, where

$$\Sigma_{j,j'}^{(il,K)} = \sum_{n} \sum_{n'}^{N} V_{jn}^{(mt,K)} G_{n,n'}^{(ts,K)} V_{n'j'}^{(mt,K)}; \quad K = L, R. \tag{21}$$

When the island comprises a single molecule positioned above, and interacting with, the metal surface atom at position $\mathbf{n}_{xy}$, Eqs. (19)-(20) reduce to the (scalar) expression for the single molecule GF

$$G^{(sm)}(E) = \left(E - \varepsilon_m - \Sigma^{(sm)}(E)\right)^{-1} \tag{22}$$

with

$$\Sigma^{(sm)}(E) = V^{(mt)} G_{\mathbf{n}_{xy},\mathbf{n}_{xy}}^{(ts)}(E) V^{(mt)}; \quad \Gamma^{(sm)}(E) = -2\,\text{Im}\left[\Sigma^{(sm)}(E)\right] \tag{23}$$

where $G_{\mathbf{n}_{xy},\mathbf{n}_{xy}}^{(ts)}$ is obtained from (16) and is independent of the position $\mathbf{n}_{xy}$. Thus, for example, the local DOS (per molecule) for a single molecule and for a molecular layer adsorbed (on site adsorption) on the simple cubic metal surface are related within our model by (using Eqs. (12), (14), (16) (for $\Delta\mathbf{n}_{xy} = 0$), (17), (18), (22) and (23))

$$\frac{\rho^{(sm)}(E)}{\rho^{(ml)}(E)} = \frac{\text{Im}\left[\left(E - \varepsilon_m - \frac{1}{4\pi^2}\int_0^{2\pi}\int_0^{2\pi} d\theta_x d\theta_y \Sigma_{\boldsymbol{\theta}_{xy}}^{(ml)}(E)\right)^{-1}\right]}{\text{Im}\left[\frac{1}{4\pi^2}\int_0^{2\pi}\int_0^{2\pi} d\theta_x d\theta_y \left(E - E_m(\boldsymbol{\theta}_{xy}) - \Sigma_{\boldsymbol{\theta}_{xy}}^{(ml)}(E)\right)^{-1}\right]} \tag{24}$$

Finally, consider adsorbed molecular layers with more complex adsorption structures. In such structures a distribution of distances between the molecules and the underlying surface atoms does not offer the simplicity of one-to-one correspondence. As long as one can identify periodicity in the adsorption structure it is possible to proceed by a straightforward generalization of the procedure described above. Eq. (4) is now generalized to

$$\left|\alpha;\theta_x\theta_y n_z\right\rangle = \frac{1}{\sqrt{N_x N_y}} \sum_{n_x=-(N_x/2)}^{(N_x/2)-1} \sum_{n_y=-(N_y/2)}^{(N_y/2)-1} e^{i\left(\theta_x n_x + \theta_y n_y\right)} \left|\alpha;n_x n_y n_z\right\rangle \tag{25}$$

where $\mathbf{n}_{xy} = (n_x, n_y)$ now stands of a unit cell and the index α stand for different sites within this cell. Note that the numbers of these sites in the adsorbate layer, $n_z = 0$, and in the underlying metal layers, $n_z \geq 1$ are in general different. We will denote them by $v^{(m)}$ and $v^{(t)}$ respectively. In this representation the metal Hamiltonian takes the form (5)-(6) where now each element in (6) is replaced by a $v^{(t)} \times v^{(t)}$ matrix. Similarly, Eqs. (9) and (10) maintain their general form, with each element replaced by a similar size matrix. Eq. (11) then becomes

$$\Gamma^{(ts)}_{\boldsymbol{\theta}_{xy}}(E) = i\left(\Sigma^{(ts)}_{\boldsymbol{\theta}_{xy}} - \Sigma^{(ts)\dagger}_{\boldsymbol{\theta}_{xy}}\right).$$

Their solution may be obtained iteratively, but is more effectively done by the renormalization group technique[19-22] The result is another representation of the metal surface GF (9), now expressed as a matrix whose elements $G^{(ts)}_{\alpha,\alpha'}(\boldsymbol{\theta}_{xy}; E)$ span the metal atoms contained within the lateral unit cell. Within the same unit cell the coupling $V^{(mt)}$ between the molecular layer and the underlying metal atoms is a rectangular matrix of order $v^{(m)} \times v^{(t)}$. With this understanding, Eq. (14) becomes an expression for the $v^{(m)} \times v^{(m)}$ self energy matrix associated with the ML molecules within the structure's unit cell. For each (α,α') element the transformation to unit cell position space is done as in Eq. (16).

The expressions derived above make it possible to compare spectral and transport properties for adsorbed molecular layers and islands, down to a single molecule. While our study focuses on the 2-dimensional interface of a 3-dimensional junction, more insight can be gained from lower dimensional system, e.g. the 1-dimensional interface between 2-dimensional leads. The corresponding GFs and SEs are similarly derived from reduced forms of the expressions given above and will be not be reproduced here.

## 4. Transmission coefficients

According to the Landauer formula,[23, 24] conduction in the linear response limit is given by $(e^2/\pi\hbar)\mathcal{T}(E_F)$ where the transmission coefficient $\mathcal{T}(E)$ is given in terms of the GF and the SE of the subsystem comprising the molecular bridge

$$\mathcal{T}(E) = \text{Tr}_{\text{bridge}}\left[\Gamma^{(L)}(E)G(E)\Gamma^{(R)}(E)(G(E))^{\dagger}\right] \quad (26)$$

Tr$_{\text{bridge}}$ stands trace over the states of the bridging system. For a finite molecular island this trace is easily evaluated in the representation of local island states $j$, using Eqs. (19) and (20). In particular, for a single molecule junction this yields

$$\mathcal{T}^{(sm)}(E) = \Gamma^{(sm,L)}(E) G^{(sm)}(E) \Gamma^{(sm,R)}(E) \left(G^{(sm)}(E)\right)^{\dagger} \quad (27)$$

For the simple molecular layer the transmission *per molecule* (denoted $\mathcal{T}^{(ml)}(E)$ below) is given by $\langle \mathbf{n}_{xy} | \Gamma^{(L)}(E) G(E) \Gamma^{(R)}(E) (G(E))^{\dagger} | \mathbf{n}_{xy} \rangle$ where $\mathbf{n}_{xy}$ corresponds the lateral position of any one molecule on the ML. This leads to[3]

$$\mathcal{T}^{(ml)}(E) = \frac{1}{4\pi^2} \int_0^{2\pi} \int_0^{2\pi} d\theta_x d\theta_y \Gamma^{(ml,L)}_{\boldsymbol{\theta}_{xy}}(E) G^{(ml)}_{\boldsymbol{\theta}_{xy}}(E) \Gamma^{(ml,R)}_{\boldsymbol{\theta}_{xy}}(E) \left(G^{(ml)}_{\boldsymbol{\theta}_{xy}}(E)\right)^{\dagger} \quad (28)$$

In the more general case (25), where each $G$ and $\Gamma$ is represented by a $\nu^{(m)} \times \nu^{(m)}$ matrix, Eq. (28), replaced by the corresponding trace over the ML unit cell, gives the ML transmission coefficient per such unit cell.

Once the transmission function (27) or (28) has been obtained, we can also compute the current through the junction using the Landauer formula,

$$I(\Phi) = \int dE \left(f_L(E) - f_R(E)\right) \mathcal{T}^{(M)}(E, \Phi) \quad (29)$$

Where $M$ stands for s single molecule (SM), a molecular island (IL) or a molecular monolayer (ML), and where, $f_L(E)$ and $f_R(E)$ are the Fermi functions of the left and right electrode, respectively.

$$f_K(E) = \left(1 + \exp\left(\frac{E - \mu + e\Phi^K}{k_B T}\right)\right)^{-1} ; \quad K = L, R, \quad (30)$$

where $\mu$ is the unbiased Fermi energy, $\Phi^K$ is the potential on the electrodes $K = L, R$, and $k_B$ and $T$ are the Boltzmann constant the temperature, respectively. In the calculation reported below we set $\Phi^L = 0$ and denote $\Phi^R = \Phi$. The latter assignment is expressed by shifting all metal sites energies on the right by $-e\Phi$. The molecular site energy is then taken $\varepsilon_m(\Phi) = \varepsilon_m(0) - Se\Phi$ where the shift parameter

---

[3] $\mathcal{T}^{(SAM)}(E)$ can be easily evaluated also in position space,

$$\mathcal{T}(E) = \sum_{\mathbf{n}_{xy}'} \sum_{\mathbf{n}_{xy}''} \sum_{\mathbf{n}_{xy}'''} \Gamma^{(ml,L)}_{\mathbf{n}_{xy},\mathbf{n}_{xy}'}(E) G^{(ml)}_{\mathbf{n}_{xy}',\mathbf{n}_{xy}''}(E) \Gamma^{(ml,R)}_{\mathbf{n}_{xy}'',\mathbf{n}_{xy}'''}(E) \left(G^{(ml)}_{\mathbf{n}_{xy}''',\mathbf{n}_{xy}}(E)\right)^{\dagger}$$ . While this expression involves infinite sums it is found that the matrix elements $G$ and $\Gamma$ fall off quickly with the distance from the diagonal, which makes these sums converge quickly.

$0 \leq S \leq 1$ reflects a particular assumption about the way by which the bias voltage falls along the molecular bridge.

## 5. Interaction parameters

The single electron tight-binding model described above depends on a number of energetic parameters that have to be estimated from experimental and computational data:

(a) The tight binding metal interaction parameter $V^{(t)}$ is chosen to give the known order, $W \sim 10 eV$, of a metallic bandwidth. In the nearest-neighbor tight-binding model $W = 4V^{(t)}d$ where $d$ is the dimensionality. Consequently $V^{(t)}$ is of the order ~1eV. Below we use $V^{(t)} = 0.82 \text{eV}$.

(b) Intermolecular interactions are estimated by computing the HOMO splitting of two molecules in a parallel configuration brought into their typical intermolecular distance on the metal surface. Using 1,4, Butanedithiol ($C_4H_{10}S_2$) aligned in parallel with intermolecular distance d=4.99Å (that characterizes a monolayer of such molecules adsorbed on Au(111) this yields[4] $V^{(m)} \sim 0.1 \text{eV}$.

(c) The molecule-metal interaction parameter, $V^{(mt)}$, is estimated from the known order of magnitude for the inverse lifetime, $\Gamma = 2\pi V^{(mt)2} \rho^{(t)} \simeq 0.01 - 1.0 eV$ for an extra electron on a molecule chemisorbed on a metal surface. Here $\rho^{(t)}$ is the metal surface density of states. The latter can be estimated from Fig. 2 to be $2\pi\rho^{(t)} \simeq 1 eV^{-1}$, which implies that $\left(V^{(mt)}\right)^2 = 0.1 - 1.0 (eV)^2$.

(d) As a model consistency check we can compute the metal contribution to the HOMO level splitting within the tight binding model using $V^{(m)} = 0.095 \text{ eV}$, and $V^{(t)} = 0.82 \text{eV}$. For $\varepsilon_m - \varepsilon_t = 0$ this calculation yields $\Delta E_{HOMO} = 0.17 eV$ and 0.19eV when $V^{(mt)} = 0.11\,\text{eV}$ and 0, respectively, while for $\varepsilon_m - \varepsilon_t = -0.25\,\text{eV}$ it yields $\Delta E_{HOMO} = 0.172\,\text{eV}$ and 0.19eV when $V^{(mt)} = 0.11\,\text{eV}$ and 0 respectively, showing a metal effect similar to that estimated from the calculations of a molecular dimmer on gold clusters.

---

[4] $\Delta E_{HOMO} \sim 2V^{(m)}$ was computed using Gaussian[25] on the HF and the LDA levels using the LANL2MB relativistic pseudo-potential basis set with added d-orbitals on the sulfur atoms. Some calculations were done on the molecular "dimer" adsorbed on gold clusters of varying sizes. The resulting splitting oscillate about $V^{(m)} = 0.1 eV$ as function of the gold cluster size.

Based on these estimates, we have used the following set of "standard" interaction and energetic parameters for our calculations

$V^{(t)}$=0.03a.u.~0.82 eV,     $V^{(m)}$=0.0035a.u.~0.095 eV,

$V^{(mt)}$=0.004a.u.~0.11 eV,    $\varepsilon_m = \varepsilon_t$ or $\varepsilon_t - 0.136$ eV.

The choice $\varepsilon_m = \varepsilon_t - 0.136\,\text{eV}$ corresponds to observed order of magnitude for good molecular hole conductors. Variations about this standard choice where made in order to gain insight on the electronic properties of this model.

## 6. Some illustrative results

Figure 2 depicts the metal surface DOS obtained from Eq. (18) using $V^{(t)} =$ 0.82 eV. Plotted are results for 1, 2 and 3 dimensional metals with the expected bandwidth of $4V^{(t)}d$ where $d$ is the dimensionality. The 1-D line is in agreement with Newns expression[26], whereas the 2-D and 3-D DOS functions display a more moderate band-edge slopes.

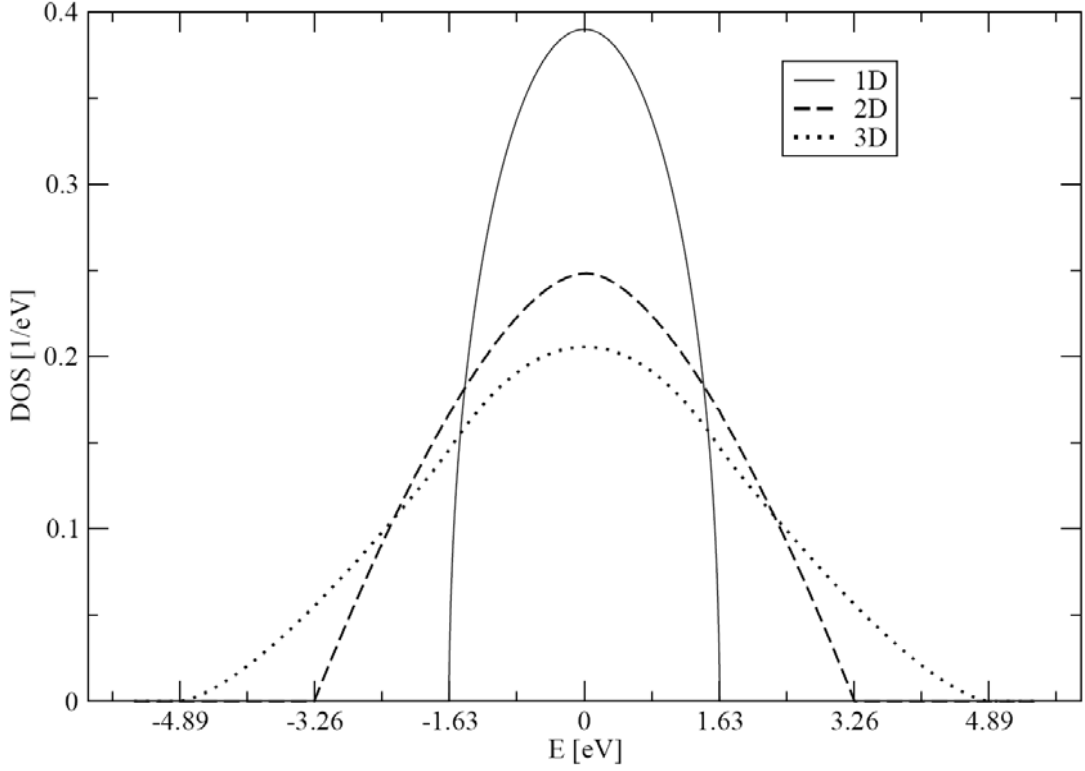

Fig.2. Surface density of states for a nearest-neighbor tight-binding model metal in 1, 2 and 3 dimensions.

The expected behavior of a single molecule and of a molecular monolayer, Eq. (24), are contrasted in Figure 3, which shows both the molecular DOS (per molecule), and the transmission coefficients, Eqs. (27),(28), of both systems when placed between two 3-dimensional metal leads. The DOS functions are scaled to coincide with the transmission functions at their maxima. The transmission and DOS functions are exactly proportional to each other in the SM case, as implied by (27) in the wide band limit. Note that the functional form of the ML density of states reflects that of the DOS of a 2-D tight-binding layer that represents the ML when removed from the metal substrates (see inset). The significant narrowing that takes place when the ML-substrate interaction is switched on indicates that direct and the through-metal intermolecular interactions are added destructively for the chosen parameters. The most important observation is the significant effect of the band-like nature of the ML density of states on the transmission coefficient which is significantly broadened relative to the corresponding SM result. Consequenctly, the SM transmission can appear significantly larger or significantly smaller than that of the ML, depending on the energy range being probed.

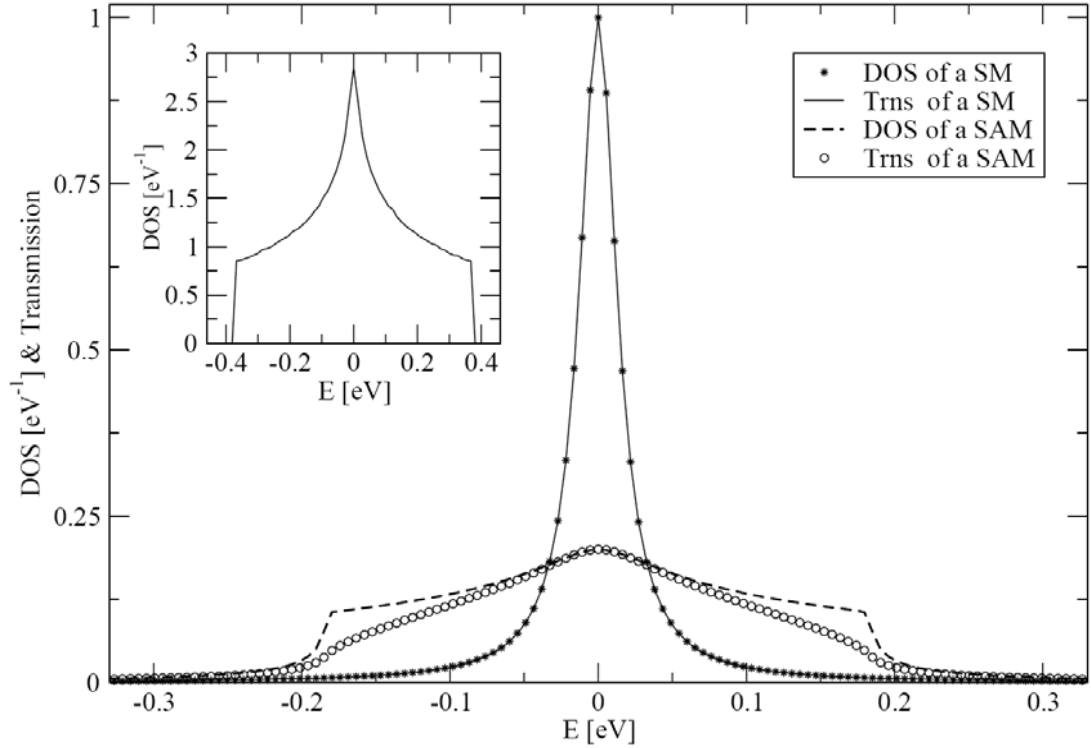

Fig.3 DOS (eV$^{-1}$) and transmission per molecule via a molecular junction, SM or ML, positioned between two identical 3D metallic electrodes. The 'standard' parameters are used with $\varepsilon_m = \varepsilon_t$ taken as the origin. The displayed DOS lines are scaled (with multiplicative factors of 0.048 and 0.088 for the SM and the ML, respectively) to coincide with the transmission at their maximal points. The inset shows the density of state of a 2-dimensional tight binding lattice with nearest neighbor coupling equal to that in the molecular layer (0.095eV).

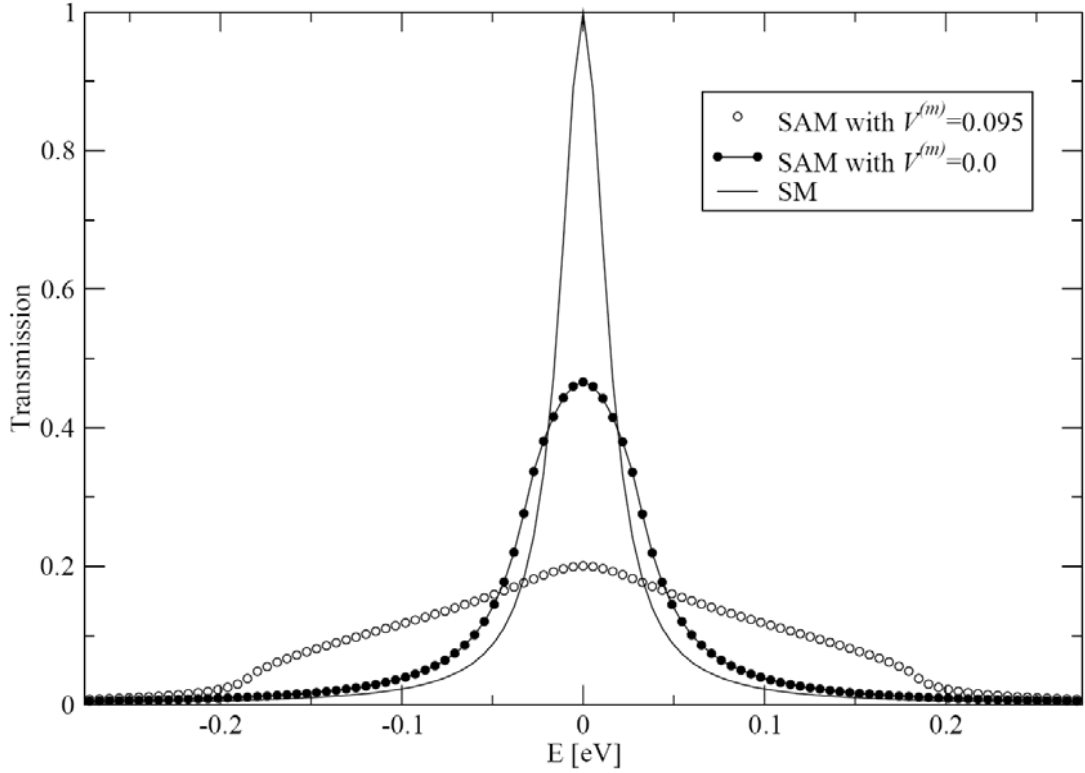

Fig. 4 Transmission (per molecule) as function of energy through a ML with zero and 'standard' intermolecular interactions. Also shown is the transmission coefficient through a single molecule (SM).

Cooperative response of the molecular layer originates both from the direct intermolecular interaction $V^{(m)}$ and from the metal induced interaction. Figure 4 shows the relative importance of these effects. It depicts the transmission per molecule through the ML with the 'standard' and zero intermolecular interaction, compared to the transmission through a single molecule. Clearly with our choice of parameter the direct intermolecular interaction plays a dominant role, even though through metal interaction has a non-negligible effect, as suggested in Ref. 7.

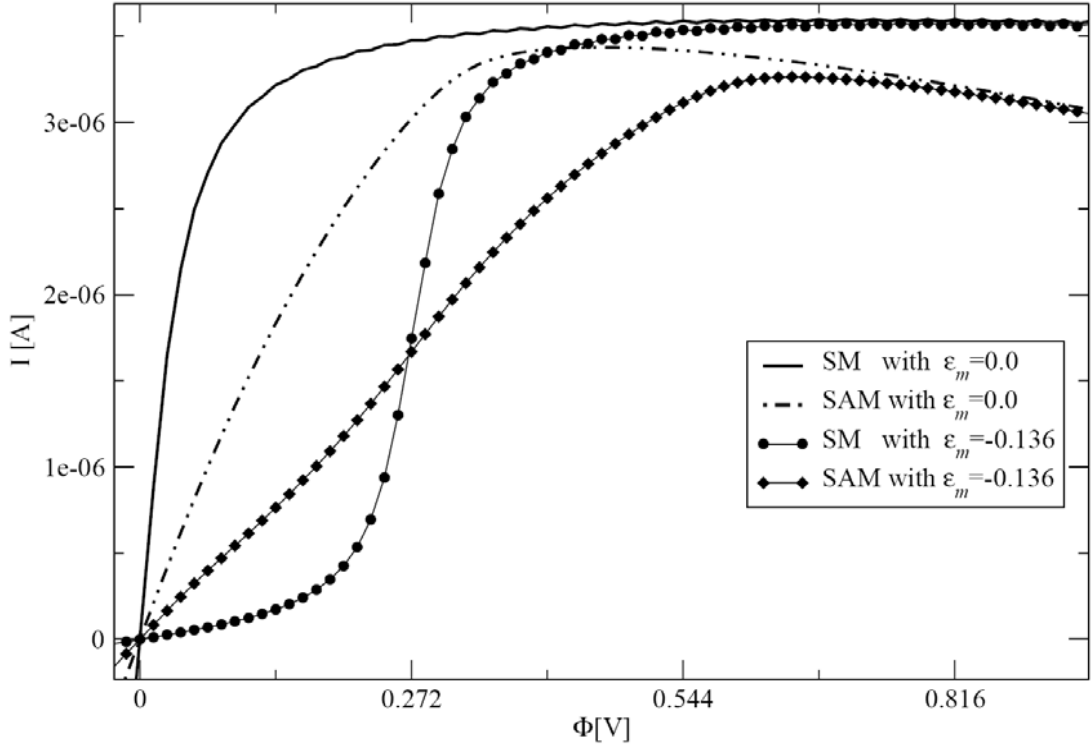

Fig 5. Current I (per molecule) as function of the bias voltage $\Phi$ via a SM and ML, with the molecular level set at $\varepsilon_m = 0$ and $\varepsilon_m = -0.136$eV. $\varepsilon_t$, taken as the energy origin is also the Fermi energy in this calculation (corresponding to a monovalent metal). In this system $I(\Phi) = I(-\Phi)$ and only the $\Phi > 0$ side is shown. For the $\varepsilon_m = -0.136$eV case, the molecular level enters the Fermi window at $\Phi = 0.272$V; in the $\varepsilon_m = 0$ case the molecular level is within this window for all $\Phi$.

The comparative plots above indicate that the main source of difference in the transport properties of a single molecule and a molecular layer is the broadened spectrum of states in the latter case. Fig. 5 shows the consequence of this broadening on the current-voltage characteristics of the junction, computed using (29) with the molecular-shift parameter $S=0.5$. When the molecular site-energy is well within the Fermi window, the SM current is higher than the current per molecule through a ML, because part of the broadened layer spectral density is outside that window. In the opposite case the molecular resonance may lie outside the Fermi window while the layer band may be broad enough to 'spill' into the window. The layer conductance will be higher in this case. Thus, in the case where $\varepsilon_m = -0.136$ eV the molecular level enters the Fermi window at $\Phi = 0.272$V. For lower bias the ML current is higher than its SM counterpart. This is reversed above this threshold where the SM spectrum is better contained within the Fermi window. The decrease in the ML current at higher voltages is a manifestation of the fact that the broadened layer spectrum starts to 'feel' the metallic band edge so that the edge of the molecular band is now falling beyond the metal band and cannot contribute to conduction. Again, level broadening arguments make it clear that such an effect will occur in a molecular layer at lower voltages than in the SM case.

Next, consider the signal scaling with the number of conducting molecules. Experimentally, such considerations pertain to two possible situations: (A) an essentially infinite molecular layer is engaged by a probe (or probes) that connect to a varying number of molecules, and (B) the junction involves molecular islands of varying sizes. For the range of interaction parameters considered, we have seen that direct intermolecular interactions dominate the cooperative response of the junction. It may therefore be expected that the per-molecule conduction in case (A) will be practically independent of the number of active molecules. On the other hand, in case (B) linear scaling is expected only beyond some characteristic island size.

Figures 6 and 7 demonstrate these types of behavior. To ease the computational effort a two-dimensional junction model is considered in which the surface and the adsorbed molecular layers are represented by 1-dimensional rows of sites. The molecules are taken to couple more strongly to one of the electrodes ("substrate") then to the other (probe). Also, the tight binding parameter for the 2-dimensional "metal" is taken to be large enough to yield a sufficiently large metal bandwidth. Specifically, the following parameters were used in the calculations presented in Figures 6 and 7: $V^{(t)} = 1.1$ eV, $V^{(m)} = 0.095$ eV, $\varepsilon_m - \varepsilon_t = -0.136$ eV, $V^{(mt1)} = 0.11$ eV (substrate), $V^{(mt2)} = 0.055$ eV (probe) and $S = 0.5$.

Figure 6(a,b) depicts the transmission coefficient and the I/Φ characteristic for case (A) while Figure 7(a,b) show similar results for case (B). Clearly, within our (physically motivated) choice of parameters case (A) is characterized by linear scaling, i.e. transport per molecule is independent of the number of active molecules, albeit different from that of an isolated molecule. In sharp contrast, case (B) shows strong dependence on the molecular island size $N$, which saturates to linear scaling behavior (and converge to the ML results) only for island exceeding $N \sim 30$ molecules in this 2-dimensional model. Interestingly, the discrete spectrum of an island comprising a small number of molecule leads to a distinct structure in the current-voltage characteristics. While this is an obvious possibility, it is usually disregarded in theoretical analysis of experimental "single molecule" junctions.

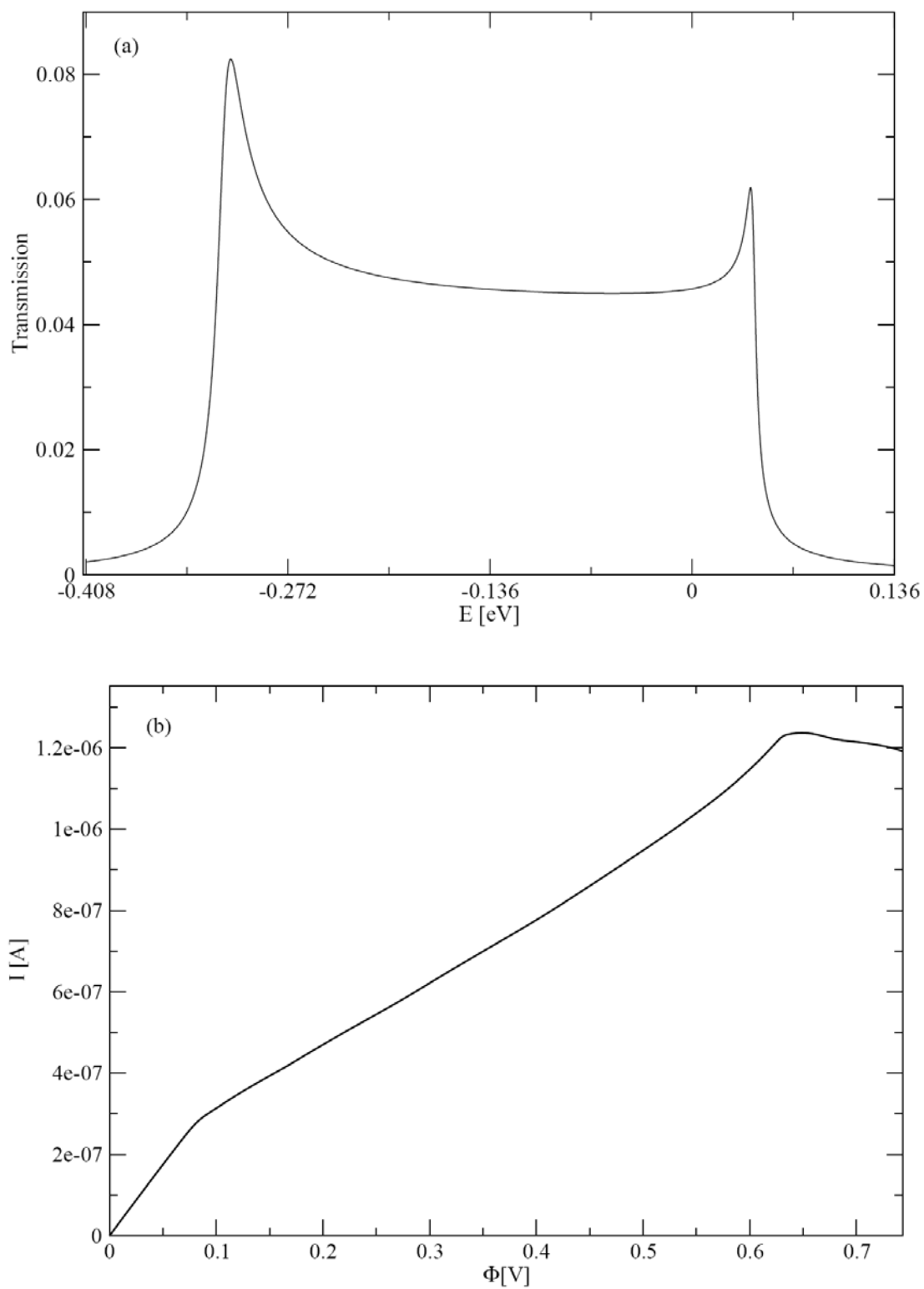

Fig. 6. The transmission coefficient (a) and the current-voltage characteristics (b) of a 2-dimensional junction in which a molecular layer adsorbed on a metallic substrate is engaged by a metallic probe that connects to different numbers of molecules. Similar plots were obtained for all numbers of molecules in the bridge (up to 30). See text for parameters.

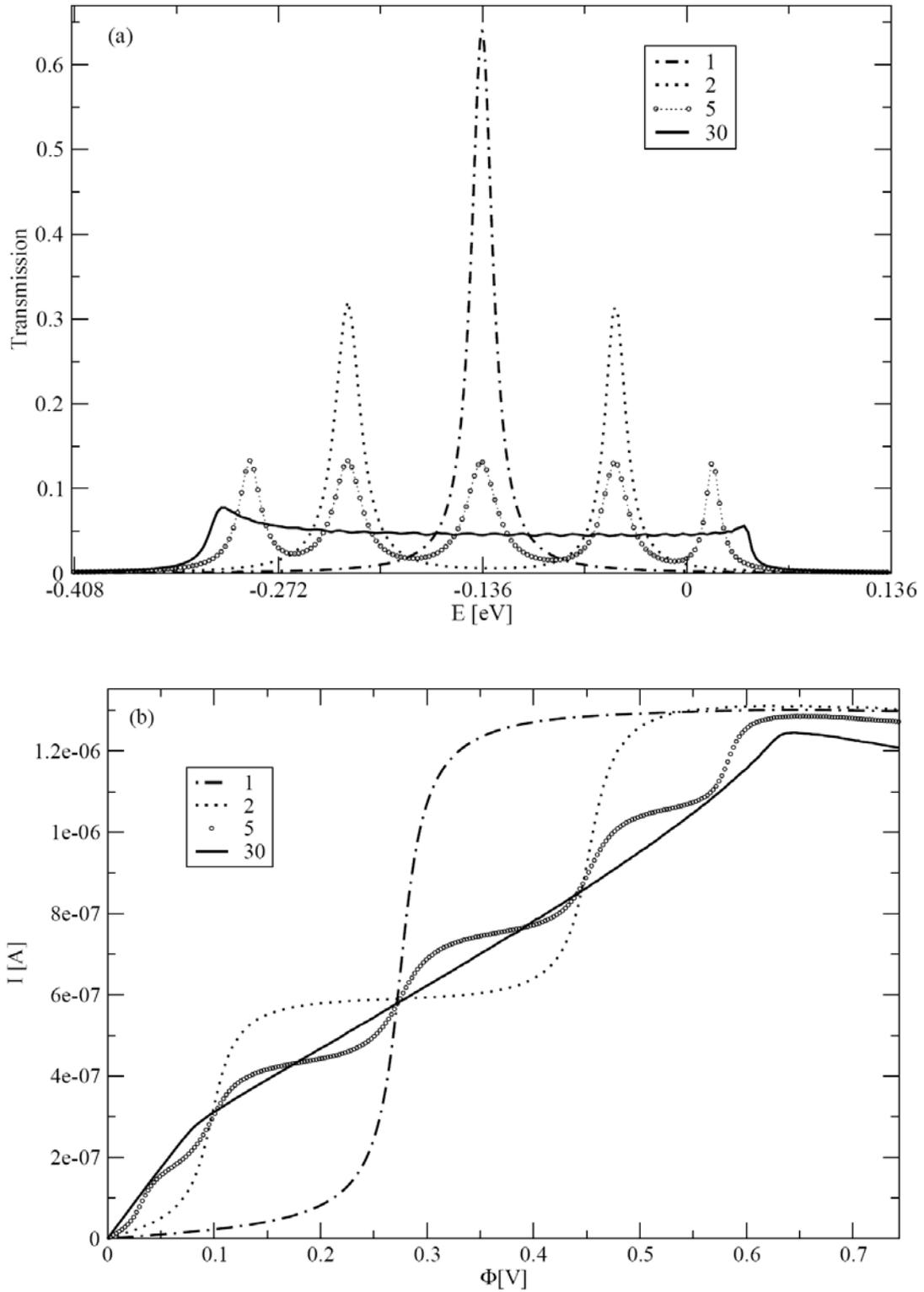

Fig. 7. The transmission coefficient (a) and the current-voltage characteristics (b) of 2-dimensional junctions comprising molecular islands of different sizes (expressed by the number of molecules N). See text for parameters.

## 7. Summary and conclusions

This paper has focused on one source of difference between the conduction properties of single molecules and the corresponding molecular islands and layers – the short range intermolecular interactions arising both from direct and through-substrate intermolecular coupling. Other important effects, in particular electrostatic interaction and charge transfer processes that depend strongly on the number of adsorbed molecules, were reviewed in the introduction. With the focus on short range interactions we have used nearest-neighbor tight-binding models for both the metal substrate and the molecular adsorbate, using parameters to fit the energetic properties of experimentally used systems. The principal source of difference between the single molecule and the molecular layer behavior arises from the different spectral properties of the molecular adsorbate, specifically the relatively narrow discrete resonance(s) associated with the single molecule adsorbate and the relatively broad 2-dimensional band structure of the molecular layer. This implies that molecular layers will conduct better in off resonance (tunneling) conduction regimes while the single molecule will conduct better (on a per-molecule basis) when resonance conditions are satisfied, i.e. where the energy of the conducting level is well within the window between the leads Fermi energies.

The individual or collective nature of molecular conduction often comes under discussion with respect to observation of apparent scaling of conduction with the number of molecules involved. We have pointed out, and demonstrated in model calculations that such observations may depend on the nature of the molecular system. Simple linear scaling of the single molecule behavior is expected in junctions where the molecular environment is fixed (e.g. a monolayer) but the electrical contact involves a varying number of molecules. However, when the junction involved molecular islands of varying sizes linear scaling is expected only beyond some limiting size. In our simplified 2-dimensional calculation, and based on our (physically motivated) choice of short range interaction parameters, we have estimated this linear size to be of the order of ~30 molecular sites. Again we emphasize that the conduction per molecule in these linear scaling regime may be quite different from that of the corresponding single molecule junction.

A layer is a generic name to many possible structures that may differ in their per-molecule conduction. Thus far we have examined systems in which the ML has a 1:1 commensurability with the metallic-surface atomic arrangement. Structures wherein the intermolecular distance in the ML differs from the unit distance in the metal reservoir possess different properties than the perfect 1:1 structure. We study the transmission through a ML with a $3a/2$ intermolecular distance, where $a$ is the metallic unit distance. Top view cartoon of the ML interaction with the metallic surface atoms in a unit cell is displayed in the inset of Fig. 8. Molecules and metallic atoms are marked with the black and gray spheres, respectively. Each molecule is interacting with four metal-atoms, for symmetric reasons there are only three types of interactions, $V_1$, $V_2$, and $V_3$. The unit cell includes 4 molecules and 9 metal-atoms; therefore, we refer to this structure as a 4:9 array. The molecule-metal interactions are assumed to depend on the distance between each molecule and its neighboring surface-atoms. Fig. 8 contrast a 1:1 layer with a 4:9 array, the parameters used in all systems are $V^{(t)} = 1.1$ eV, $V^{(m)} = 0.095$ eV, $\varepsilon_m - \varepsilon_t = -0.136$ eV, where the different molecule-metal interactions are specified within the figure.

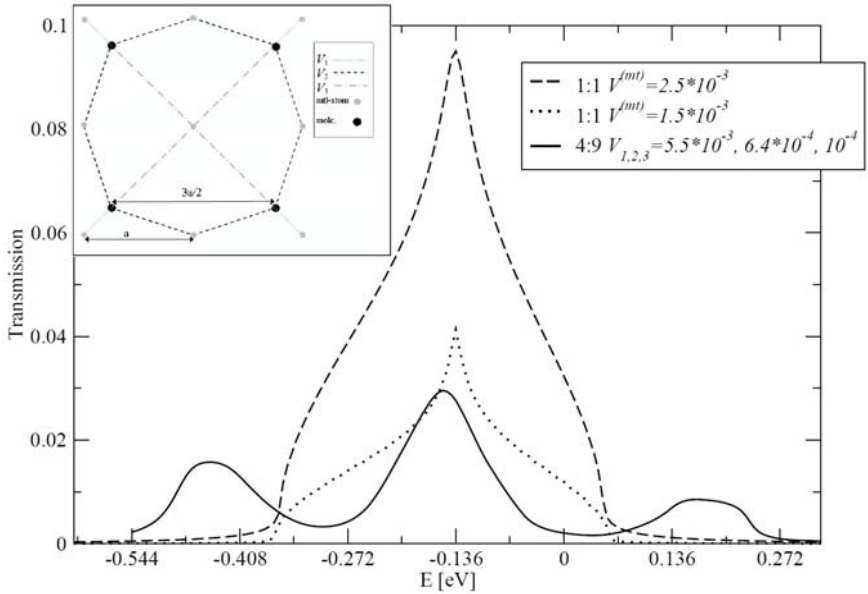

Fig. 8. Transmission functions via 1:1 vs. 4:9 ML arrangement packed between two equal 3D metals. The inset is a top view cartoon displaying the 4:9 array unit cell and the molecules-metallic-surface-atoms interactions $V_{1,2,3}$.

The three peaks in the 4:9 transmission function may be explained by the effective interactions between the ML molecules. As we established above the ML may be approximated by a 2D bulk model. However, in this case the 2D bulk needs to be represented by an interaction-alteration nearest-neighbor model, where each molecule has the same site energy and 'feels' different interactions. This model gives rise a three band electronic structure, whereas, for relatively small difference between the interactions the bands start to overlap giving rise to the peaks observed in the transmission plot.

Finally, recall the interesting observation of negative differential conduction in Fig. 5. This behavior appears in a junction comprising a molecular layer in the high bias regime, reflecting a situation where the broadened molecular band 'explores' the edge of the metallic band structure. This extreme situation for metallic molecular junctions is more common in junctions involving semiconductor leads, and its implication for the conduction properties of single molecules and molecular layers will be discussed elsewhere.

## Acknowledgements

We thank David Cahen and Ron Naaman for many helpful discussions. This research was supported by grants from the "Bikura" track of the Israel Science Foundation, the USA-Israel Binational Science Foundation, the Germany-Israel Foundation and the Schmidt Minerva Center for Supramolecular Chemistry.